\documentstyle[multicol,aps,epsf]{revtex}
\begin{document}
\title{Infinite-Order Percolation and Giant Fluctuations 
in a Protein Interaction Network}  
\author{J. Kim$^1$, P. L. Krapivsky$^2$, B. Kahng$^1$, and S. Redner$^2$}
\address{$^1$School of Physics and Center for Theoretical Physics,  Seoul
  National University, Seoul 151-747, Korea}
\address{$^2$Center for BioDynamics, Center for Polymer Studies, 
and Department of Physics, Boston University, Boston, MA, 02215}

\maketitle
\begin{abstract}
  We investigate a model protein interaction network whose links represent
  interactions between individual proteins.  This network evolves by the
  functional duplication of proteins, supplemented by random link addition to
  account for mutations.  When link addition is dominant, an infinite-order
  percolation transition arises as a function of the addition rate.  In the
  opposite limit of high duplication rate, the network exhibits giant
  structural fluctuations in different realizations.  For
  biologically-relevant growth rates, the node degree distribution has an
  algebraic tail with a peculiar rate dependence for the associated exponent.

\smallskip\noindent{PACS numbers: 02.50.Cw, 05.40.-a, 05.50.+q, 87.18.Sn}
\end{abstract}
\begin{multicols}{2}
\narrowtext
  
Inter-protein interactions underlie the performance of vital biological
functions.  Organisms with sequenced genomes, such as the yeast {\it S.
  cerevisiae}\cite{96}, provide important test beds for analyzing protein
interaction networks\cite{uetz}.  The number of interactions per protein of
{\it S.  cerevisiae} follows a power-law \cite{ito,wagner,jeong}, a feature
common to many complex networks, such as the Internet, the world-wide web,
and metabolic networks\cite{review}.  Similar behavior is exhibited by
protein interaction networks of various bacteria\cite{bacterium}.  Based on
the observational data, simple proteome growth models have recently been
formulated to account for the evolution of this interaction
network\cite{slanina,vazquez,sole,SSS}, where proteins are viewed as the
nodes of a graph and links connect functionally related proteins.

In this work, we determine the structure of a minimal protein interaction
network model that evolves by the biologically-inspired processes of protein
duplication and subsequent mutation.  That is, the functionality of a
duplicate protein is similar, but not identical, to the original and can
gradually evolve with time due to mutations\cite{wagner}.  Within a rate
equation approach\cite{krl,kr}, we show that: (i) the system undergoes an
infinite-order percolation transition as a function of mutation rate, with a
rate-dependent power-law cluster-size distribution everywhere below the
threshold, (ii) there are giant fluctuations in network structure and no
self-averaging for large duplication rate, and (iii) the degree distribution
has an algebraic tail with a peculiar rate-dependent exponent when the
duplication and mutation rates have biologically realistic values.  Some
aspects of this last result were recently seen \cite{sole,SSS}.

\begin{figure}
  \narrowtext \epsfxsize=2.9in \hskip 0.2in
\epsfbox{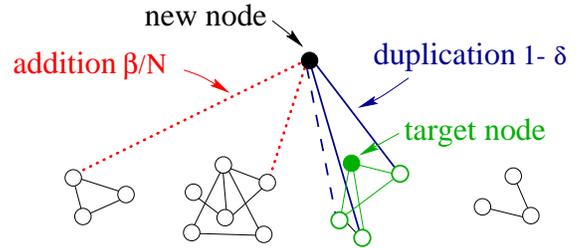} \vskip 0.1in
\caption{Growth steps of the protein interaction network: The new node 
  duplicates 2 out of the 3 links between the target node
  (shaded) and its neighbors.  Each successful duplication occurs with
  probability $1-\delta$ (solid lines).  The new node also attaches to any
  other network node with probability $\beta/N$ (dotted lines).  Thus 3
  previously disconnected clusters are joined by the complete event.
\label{model}}
\end{figure}

In the model, nodes are added sequentially and the new node duplicates a
randomly chosen pre-existing ``target'' node, {\it viz.}, the new node links
to each of the neighbors of the target with probability $1-\delta$; each new
node also links to any previous node with probability $\beta/N$, where $N$ is
the current total number of nodes (Fig.~\ref{model}).  Thus an arbitrary
number of clusters can merge when a single node is introduced.  As we now
discuss, this unusual dynamics appears to be responsible for the
unconventional percolation properties of this network in the limit of zero
duplication rate but finite mutation rate ($\delta=0$, $\beta>0$).

Let $C_s(N)$ be the expected number of clusters of size $s\geq 1$.  This cluster
size distribution obeys the rate equation
\begin{equation}
\label{Cs}
{d C_s\over dN}=-\beta\,{sC_s\over N}
+\sum_{n=0}^\infty {\beta^n\over n!}\,e^{-\beta}
\sum_{s_1\cdots s_n}\prod_{j=1}^n{s_jC_{s_j}\over N},
\end{equation}
where the sum is over all $s_1\geq 1, \ldots, s_n\geq 1$ such that
$s_1+\cdots +s_n+1=s$.  The first term on the right-hand side of
Eq.~(\ref{Cs}) accounts for the loss of $C_s$ due to the linking of a cluster
of size $s$ with the newly-introduced node.  The gain term accounts for all
merging processes of $n$ initially separated clusters whose total size is
$s-1$.

Solving for the first few $C_s(N)$, we see that they are all proportional to
$N$.  Thus writing $C_s(N)=Nc_s$, and introducing the generating function
$g(z)=\sum_{s\geq 1} sc_s\,e^{sz}$, Eq.~(\ref{Cs}) becomes
\begin{equation}
\label{g}
g=-\beta g'+(1+\beta g')\,e^{z+\beta(g-1)},
\end{equation}
where $g'=dg/dz$.  To detect the percolation transition, we use the fact that
$g(0)=\sum sc_s$ is the fraction of nodes within finite clusters.  Thus the
size of the infinite cluster (the giant component) is $NG=N(1-g(0))$.
Suppose that we are in the non-percolating phase; this means that $g(0)=1$.
In this regime, the average cluster size equals $\langle s\rangle=\sum s^2
c_s=g'(0)$.  To determine $g'(0)$, we substitute the expansion
$g(z)=1+zg'(0)+\ldots$ into Eq.~(\ref{g}) and take the $z\to 0$ limit.  This
yields a quadratic equation for $g'(0)$ with solution
\begin{equation}
\label{sav}
g'(0)=\langle s\rangle={1-2\beta-\sqrt{1-4\beta}\over 2\beta^2}.
\end{equation}
This has a real solution only for $\beta\leq {1/4}$, thus identifying the
percolation threshold as $\beta_c={1/ 4}$.  For $\beta>\beta_c$, we express
$g'(0)$ in terms of the size of the giant component by setting $z=0$ in
Eq.~(\ref{g}) to give
\begin{equation}
\label{sav1}
g'(0)={e^{-\beta G}+G-1\over \beta\left(1-e^{-\beta G}\right)}.
\end{equation}
When $\beta\to\beta_c$, we use $G\to 0$ to simplify Eq.~(\ref{sav1}) and find
$\langle s\rangle\to (1-\beta_c)\beta_c^{-2}=12$.  On the other hand,
Eq.~(\ref{sav}) shows that $\langle s\rangle\to 4$ when $\beta\to\beta_c$
from below.  Thus the average size of the {\em finite} clusters jumps
discontinuously from 4 to 12 as $\beta$ passes through $\beta_c={1\over 4}$.

The cluster size distribution $c_s$ exhibits distinct behaviors below, at,
and above the percolation transition.  For $\beta<\beta_c$, the asymptotic
behavior of $c_s$ can be read off from the behavior of the generating
function as $z\to 0$.  If $c_s$ has the power-law behavior
\begin{equation}
\label{asym}
c_s\sim B\, s^{-\tau}\quad{\rm as} \quad s\to\infty, 
\end{equation}
then the corresponding generating function $g(z)$ has the following small-$z$
expansion
\begin{eqnarray}
\label{expan} 
g(z)=1+g'(0)\,z+B\Gamma(2-\tau)\,(-z)^{\tau-2}+\ldots.
\end{eqnarray}
The regular terms are needed to reproduce the known zeroth and first
derivatives of the generating function, while the asymptotic behavior is
controlled by the dominant singular term $(-z)^{\tau-2}$.  Higher-order
regular terms are asymptotically irrelevant.  Substituting this expansion
into Eq.~(\ref{g}) we find that the dominant terms are of the order of
$(-z)^{\tau-3}$.  Balancing all contributions of this order gives
\begin{equation}
\label{tau}
\tau=1+{2\over 1-\sqrt{1-4\beta}}.
\end{equation}
Intriguingly, a power-law cluster size distribution with a non-universal
exponent arises for {\em all} $\beta<\beta_c$.  In contrast to ordinary
critical phenomena, the entire range $\beta<\beta_c$ is critical.

The power-law tail implies that the size of the largest cluster $s_{\rm max}$
grows as a power law of the system size.  From the extreme statistics
criterion \hbox{$\sum_{s\geq s_{\rm max}}N\,c_s=1$} and the asymptotics of
Eq.~(\ref{asym}), we find \hbox{$s_{\rm max}\propto N^{1/(\tau-1)}$}, or
\hbox{$s_{\rm max}\propto N^{{1\over 2}-\sqrt{\beta_c-\beta}}$}.  In
contrast, for conventional percolation below threshold, the largest cluster
has size $s_{\rm max}\propto\ln N$, reflecting the exponential tail of the
cluster size distribution \cite{percol}.

At the transition, Eq.~(\ref{tau}) gives $\tau=3$.  However, the naive
asymptotics $c_s\propto s^{-3}$ cannot be correct as it implies that $g'(0)$
diverges.  Similarly, we cannot expand the generating function as in
Eq.~(\ref{expan}) with $\tau=3$, since the singular term $\Gamma(-1) \times
(-z)$ has an infinite prefactor.  As in other situations where the order of a
singular term coincides with a regular term, we anticipate a logarithmic
correction.  Thus consider the modified expansion $g(z)=1+4z+z\,
u(z)+\ldots$, where $u(z)$ vanishes slower than any power of $z$, as $z\to
0$.  Substituting this into Eq.~(\ref{g}), setting $\beta=\beta_c$, and
equating singular terms yields $(8+u)\,z\,u'+u^2=0$.  Solving this
differential equation asymptotically we obtain the leading behavior $u\approx
{8/\ln(-z)}$; this indeed vanishes slower than any power of $z$ for $z\to 0$.
Substituting this form for $u(z)$ in the modified expansion for $g(z)$ and
inverting yields
\begin{equation}
\label{asym1}
c_s\sim {8\over s^3\,(\ln s)^2}\quad{\rm as} \quad s\to\infty.
\end{equation}
Thus exactly at the transition, the cluster size distribution acquires a
logarithmic correction.  This result also implies that the size of the
largest component scales as $s_{\rm max}\propto N^{1/2}/\ln N$.

Above the percolation transition, both $g(0)=1-G$ and $g'(0)$
(Eq.~(\ref{sav1})) are finite, so that the expansion for $g(z)$ has the form
$g(z)=1-G+g'(0)\,z+\ldots$.  Substituting this into Eq.~(\ref{g}) one can
show that: (i) the full expansion of $g(z)$ is regular in $z$, and (ii) the
generating function diverges at $z_*=1/s_*$.  This latter fact implies that
$c_s\propto e^{-s/s_*}$ as $s\to\infty$.  The location of the singularity is
determined by the condition $e^{z+\beta(g-1)}=1$.  This gives $s_*\to {16/G}$
as $\beta\to \beta_c$.  Realistic protein interaction networks are always
above the percolation transition, {\it e.g.}, for yeast the giant component
includes 54\% of all nodes and 68\% of the links of the system\cite{ito};
thus a giant component always exists and the cluster-size distribution has an
exponential tail.

The size of the giant component $G(\beta)$ is obtained by solving
Eq.~(\ref{g}) near $z=0$.  A lengthy analysis\cite{future} shows that near
the percolation threshold:
\begin{equation}
\label{giant}
G(\beta)\propto \exp\left(-{\pi\over\sqrt{4\beta-1}}\right),
\end{equation}
so that all derivatives of $G(\beta)$ vanish as $\beta\to \beta_c$.  Thus the
transition is of infinite order.  Similar behavior has been recently observed
\cite{clusters,sam,bernard,kr} for several growing network models where
single nodes and links were introduced independently.  This generic growth
mechanism seems to give rise to fundamentally new percolation phenomena.

We now examine the complementary limit of no mutations ($\beta=0$) and show
that individual realizations of the evolution lead to widely differing
results.  Consider first the limit of deterministic duplication of $\delta=0$
where all the links of the duplicated protein are completed.  There is still
a stochastic element in this growth, as the node to be duplicated is chosen
randomly.  When $\delta=0$, the rate equation approach
[Eqs.~(\ref{Nk})--(\ref{ab}) below] predicts that the degree distribution
$N_k$ (defined as the number of nodes that are linked to $k$ other nodes) is
given by $N_k=2(1-2/N)^{k-1}$.  

\begin{figure}
  \narrowtext \hskip 0.1in\epsfxsize=2.9in
\epsfbox{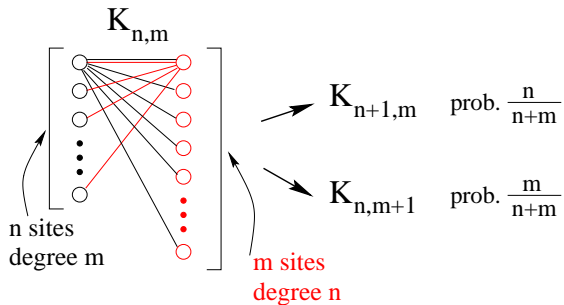} 
\caption{Evolution of the complete bipartite graph $K_{m,n}$ after 
  one deterministic duplication event.  Only the links emanating from the top
  nodes of each component are shown.
\label{Kmn}}
\end{figure}

However, this ``solution'' does not correspond to the outcome of any single
realization of the duplication process.  To appreciate this, consider the
simple and generic initial state of two nodes that are joined by a single
link.  We denote this graph as $K_{1,1}$, following the graph theoretic
terminology\cite{graph} that $K_{n,m}$ denotes a complete bipartite graph in
which every node in the subgraph of size $n$ is linked to every node in the
subgraph of size $m$.  Duplicating one of the nodes in $K_{1,1}$ gives
$K_{2,1}$ or $K_{1,2}$, equiprobably.  By continuing to duplicate nodes, one
finds that at every stage the network always remains a complete bipartite
graph, say $K_{k,N-k}$, and that every value of $k=1,\ldots,N-1$ occurs with
equal probability (Fig.~\ref{Kmn}).  Thus the degree distribution remains
singular -- it is always the sum of two delta functions!

For fixed $N$, we average over all
realizations of the evolution to obtain the {\em average} degree distribution
\begin{equation}
\label{Nkav}
\langle N_k\rangle=2\left(1-{k-1\over N-1}\right).
\end{equation}
Computing $\langle N_k\rangle$ for other generic initial conditions, {\it
  e.g.}, complete $m$-partite graphs and ring graphs\cite{future}, we find
that the initial condition dependence persists throughout the evolution.
More importantly, self-averaging breaks down: different realizations of the
growth lead to statistically distinguishable networks.  Similar giant
fluctuations arise in the general case of imperfect duplication where
$\beta=0$ and $\delta>0$ \cite{future}.  To illustrate the origin of these
macroscopic fluctuations, consider the network growth in the limit $\delta\ll
1$.  The probability that the first few duplication steps are complete (all
eligible links are created) is close to one.  For this initial development,
the degrees of each node increase and the probability to create isolated
nodes becomes very small as the network grows.  On the other hand, if the
first duplication event was totally incomplete, an isolated node would be
created.  The creation of isolated nodes necessarily leads to more isolated
nodes but subsequent duplication events.  Thus the number of isolated nodes
is a non-self-averaging quantity.  In a similar fashion, the number of nodes
of degree $k$ for any finite $k>0$ is also non-self-averaging.
 
Finally, we investigate to the evolution of the network when both incomplete
duplication and mutation occur ($\delta<1$ and $\beta>0$).  Let us first
determine the average node degree of the network, ${\cal D}$, for such
general rates.  In each growth step, the average number of links $L$
increases by $\beta+(1-\delta){\cal D}$.  Therefore,
$L=[\beta+(1-\delta){\cal D}]N$.  Combining this with ${\cal D}=2L/N$
gives\cite{vazquez,sole}
\begin{equation}
\label{d}
{\cal D}={2\beta\over 2\delta-1},
\end{equation}
a result that applies only when $\delta>\delta_c=1/2$.  Below this threshold,
the number of links grows as
\begin{equation}
\label{L}
{d L\over dN}=\beta+2(1-\delta)\,{L\over N},
\end{equation}
and combining with ${\cal D}(N)=2L(N)/N$, we find
\begin{equation}
\label{cases}
{\cal D}(N)=\cases{{\rm finite}& $\delta>1/2$,\cr
            \beta \ln N & $\delta=1/2$,\cr
            {\rm const.}\times N^{1-2\delta}& $\delta<1/2$.}
\end{equation}
Without mutation ($\beta=0$) the average node degree always scales as
$N^{1-2\delta}$, so that a realistic finite average degree is recovered {\em
  only} when $\delta=1/2$.  Thus mutations play a constructive role, as a
finite average degree arises for any duplication rate $\delta>1/2$.
 
We now consider this case of $\delta>1/2$ and $\beta>0$ and apply the rate
equation approach \cite{krl,kr} to study the degree distribution $N_k(N)$.
The degree $k$ of a node increases by one at a rate $A_k=(1-\delta)k+\beta$.
The first term arises because of the contribution from duplication, while
mutation leads to the $k$-independent contribution.  The rate equations for
the degree distribution are therefore
\begin{equation}
\label{Nk}
{d N_k\over dN}={A_{k-1} N_{k-1}-A_k N_k\over N}+G_k.
\end{equation}
The first two terms account for processes in which the node degree increases
by one.  The source term $G_k$ describes the introduction of a new node of
$k$ links, with $a$ of these links created by duplication and $b=k-a$ created
by mutation.  The probability of the former is $\sum_{s\geq a} n_s{s\choose
  a}(1-\delta)^a \delta^{s-a}$, where $n_s=N_s/N$ is the probability that a
node of degree $s$ is chosen for duplication, while the probability of the
latter is $\beta^b\,e^{-\beta}/b!$.  Since duplication and random attachment
are independent processes, the source term is
\begin{equation}
\label{ab}
G_k=\sum_{a+b=k}\sum_{s=a}^\infty 
n_s\,{s\choose a}(1-\delta)^a \delta^{s-a}\,{\beta^b\over b!}\,e^{-\beta}.
\end{equation}

{}From Eq.~(\ref{Nk}), the $N_k$ grow linearly with $N$. 
Substituting $N_k(N)=N\,n_k$ in the rate equations yields
\begin{equation}
\label{nk}
\left(k+{\beta+1\over 1-\delta}\right) n_k=
\left(k-1+{\beta\over 1-\delta}\right) n_{k-1}+{G_k\over 1-\delta}.
\end{equation}
Since $G_k$ depends on $n_s$ for all $s\geq k$, the above equation is not a
recursion.  However, for large $k$, we can reduce it to a recursion by simple
approximations.  As $k\to\infty$, the main contribution to the sum in
Eq.~(\ref{ab}) arises when $b$ is small, so that $a$ is close to $k$, and the
summand is sharply peaked around $s\approx k/(1-\delta)$.  This simplifies
the sum, as we may replace the lower limit by $s=k$, and $n_s$ by its value
at $s=k/(1-\delta)$.  Further, if $n_k$ decays as $k^{-\gamma}$, we write
$n_s=(1-\delta)^\gamma n_k$ and simplify $G_k$ to
\begin{eqnarray}
\label{Gks}
G_k &\approx& (1-\delta)^\gamma\,n_k \sum_{s=k}^\infty
{s\choose k}(1-\delta)^k \delta^{s-k}\, 
\sum_{b=0}^\infty{\beta^b\over b!}\,e^{-\beta}\nonumber \\
  &=&(1-\delta)^{\gamma-1}n_k,
\end{eqnarray}
since the former binomial sum equals $(1-\delta)^{-1}$.  

\begin{figure}
  \narrowtext \hskip 0.1in\epsfxsize=2.9in
\epsfbox{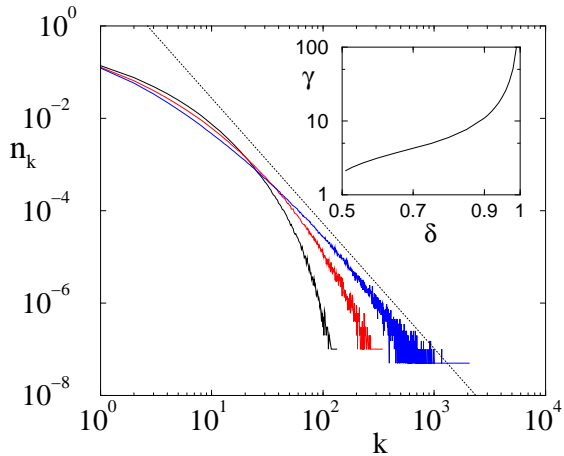} \vskip 0.1in
\caption{Degree distribution $n_k$ versus $k$ for the protein interaction
  network with $\delta=0.53$ and $\beta=0.06$.  Shown is the distribution for
  $N=10^3$, $10^4$, and $10^6$ (bottom to top), with $10^4$, $10^3$, and $20$
  realizations respectively.  A straight line (dotted) of the predicted slope
  of $-2.37$ is shown for visual reference.  The inset shows the degree
  distribution exponent $\gamma$ as a function of $\delta$ from the numerical
  solution of Eq.~(\ref{gamma}).
\label{dd}}
\end{figure}

Thus for $k\to\infty$, Eq.~(\ref{nk}) reduces to a recursion relation, from
which we deduce that $n_k$ has the power-law behavior $\sim k^{-\gamma}$,
with $\gamma$ determined from the relation
\begin{equation}
\label{gamma}
\gamma(\delta)=1+{1\over 1-\delta}-(1-\delta)^{\gamma-2}. 
\end{equation}
Notice that the replacement of $n_s$ by $(1-\delta)^\gamma n_k$ is valid only
asymptotically.  This explains the slow convergence of the degree
distribution to the predicted power law form (Fig.~\ref{dd}).  Intriguingly,
the exponent $\gamma(\delta)$ is independent of the mutation rate $\beta$
\cite{extra}.  Nevertheless, the presence of mutations ($\beta>0$) is vital
to suppress the non-self-averaging as the network evolves and thus make
possible a smooth degree distribution.  If we adopt $\delta=0.53$, as
suggested by observations\cite{wagner}, we obtain $\gamma=2.373\ldots$,
compared to the numerical simulation result of $\gamma=2.5\pm
0.1$\cite{sole}.

In summary, network growth by duplication and mutation leads to rich behavior
with an infinite-order percolation transition and no self-averaging in the
absence of mutations.  Without mutation, different realizations of the
network lead to drastically different outcomes and each outcome is itself
singular.  Mutations are needed to form networks that are statistically
similar to observed protein interaction networks.  Thus mutations seem to
play a constructive role in forming robust networks whose functioning
realizes the primary purpose of mutations.

We thank NSF grant DMR9978902, KOSEF grant 2002-2-11200-002-3 in the BRP
program, and a travel grant from the BK21 project.

\end{multicols}

\begin{thebibliography}{99}

\bibitem{96} 
   A.~Goffeau {\it et al.}, Science {\bf 274}, 546 (1996).

\bibitem{uetz} 
   P.~L.~Uetz {\it et al.}, Nature {\bf 403}, 623 (2000);
   E.~M.~Marcotte {\it et al.}, Nature {\bf 402}, 83 (1999);
   A.~J.~Enright {\it et al.}, Nature {\bf 402}, 86 (1999). 

\bibitem{ito} 
   T.~Ito {\it et al.}, Proc. Natl. Acad. Sci. USA {\bf 97}, 
   1143 (2000); $\it ibid$ {\bf 98}, 4569 (2001). 

\bibitem{wagner}
   A.~Wagner, Mol.\ Biol.\ Evol.\ {\bf 18}, 1283 (2001).  

\bibitem{jeong} 
   H.~Jeong {\it et al.}, Nature {\bf 411}, 41 (2001).
   
\bibitem{review} 
   For reviews, see S.~H.~Strogatz, Nature {\bf 410}, 268 (2001);
   R.~Albert and A.-L.~Barab\'asi, Rev.\ Mod.\ Phys.\ {\bf 74}, 47 (2002);
   S.~N.~Dorogovtsev and J.~F.~F.~Mendes, Adv.\ Phys. {\bf 51}, 1079 (2002).

\bibitem{bacterium}  
   J.-C.~Rain {\it et al.}, Nature {\bf 409}, 211 (2001).
   
\bibitem{slanina} 
   A similar network growth mechanism was proposed earlier in
   the context of biological evolution: F. Slanina and M. Kotrla, 
   Phys.\ Rev.\ E {\bf 62}, 6170 (2000).
   
\bibitem{vazquez} 
   A.~Vazquez, A.~Flammini, A.~Maritan, and A.~Vespignani,
   {\it cond-mat}/0108043.

\bibitem{sole} 
   R.~V.~Sol\'e, R.~Pastor-Satorras, E.~D.~Smith, and T.~Kepler,
   Adv.\ Complex\ Syst. {\bf 5}, 43 (2002). 
   
\bibitem{SSS} 
   R.~Pastor-Satorras, E.~D.~Smith, and R.~V.~Sol\'e, preprint.   

\bibitem{krl} 
   P.~L.~Krapivsky, S.~Redner, and F.~Leyvraz, Phys.\ Rev.\ 
   Lett.\ {\bf 85}, 4629 (2000); P.~L.~Krapivsky and S.~Redner, Phys.\ Rev.\ 
   E {\bf 63}, 066123 (2001).

\bibitem{kr} 
   P.~L.~Krapivsky and S.~Redner, {\it cs.GL}/0206011.

\bibitem{percol}
   D.~Stauffer and A.~Aharony, 
   {\em Introduction to Percolation Theory} (Taylor and Francis, 
   London, 1992).

\bibitem{future} 
   J.~Kim, P.~L.~Krapivsky, B.~Kahng, and S.~Redner, (in preparation).
   
\bibitem{clusters} 
   D.~S.~Callaway {\it et al.}, Phys.\ Rev.\ E {\bf 64}, 041902 (2001). 

\bibitem{sam} 
   S.~N.~Dorogovtsev, J.~F.~F.~Mendes,  and A.~N.~Samukhin,
   Phys.\ Rev.\ E {\bf 64}, 066110 (2001).

\bibitem{bernard} 
   M.~Bauer and D.~Bernard, {\it cond-mat}/0203232.

\bibitem{graph} 
   B.~Bollob\'as, {\em Modern Graph Theory} (Springer, New York, 1998).
   
 \bibitem{extra} Ref.~\cite{SSS} used an approximation to the rate equations
   that apply only as $\delta\to0$ and obtained a different form for the
   exponent $\gamma$ than that predicted in Eq.~(\ref{gamma}).
 
\end{thebibliography}
\end{document}